\documentclass[a4paper,11pt]{article}
\usepackage{epsfig,sint,cite,mathbbol,graphicx}
\usepackage{subfig,color}
\usepackage{bm}

\begin{document}
\begin{titlepage}
\begin{flushright}
MKPH-T-12-12 \\ HIM-2012-07
\end{flushright}

\renewcommand{\thefootnote}{\fnsymbol{footnote}}

\vskip 0.5 cm
\begin{center}
%%  {\Large\bf Towards a benchmark lattice calculation of the nucleon
%%  axial charge \\[0.5ex]}
  {\Large\bf Improved interpolating fields for hadrons at non-zero momentum
  \\[0.5ex]}
\end{center}
\vskip 1.0cm
\begin{center}
{\large
  M.\,Della Morte$^{a,b}$,
  B.\,J\"ager$^{a,b}$,
  T.~D.\, Rae$^a$,
  H.\,Wittig$^{a,b}$%\footnote{morte@kph.uni-mainz.de}
}
\vskip 1.0cm
$^{a}$\,PRISMA Cluster of Excellence, Institut f\"ur Kernphysik, Becher Weg 45, University of Mainz,
  D-55099 Mainz, Germany  
\vskip 0.3cm
$^{b}$\,Helmholtz Institute Mainz, University of Mainz, D-55099 Mainz,
  Germany 
\vskip 2.5cm
{\bf Abstract}
\vskip 1.0ex
\end{center}

\renewcommand{\thefootnote}{\arabic{footnote}}

\noindent
We generalise Gaussian/Wuppertal smearing in order to produce
non-spherical wave functions. We show that we can achieve a reduction
in the noise-to-signal ratio for correlation functions of certain hadrons at non-zero momentum,
while at the same time preserving a good projection on the ground state.
\vfill

\eject

\vfill
\eject

\end{titlepage}

\setcounter{footnote}{0}

\section{Introduction}

In lattice QCD, masses, energies and transition matrix elements of asymptotic states with
 given  quantum numbers are typically extracted from the Euclidean
time dependence of suitable correlation functions. The contribution of the lowest state
can be separated from those of other states by inserting the source fields at large-enough
time distances. The associated statistical error can be estimated from the spectral
properties of the theory~\cite{Parisi,Lepage} and usually grows exponentially with the time
separation\footnote{One exception is the pion channel (at zero momentum), for which the signal-to-noise ratio
is asymptotically constant as a function of the source-sink separation.}. 
In practice, this makes it very difficult to find a window where statistical and
systematic errors are both under control. The use of smeared sources (i.e. interpolating fields with 
small overlaps on the excited states) ameliorates the situation, as the contribution from the
 lowest state is then expected to saturate the correlation function already at small
Euclidean separations.

\begin{figure}[h!]
\centering
\includegraphics[width=0.8\linewidth]{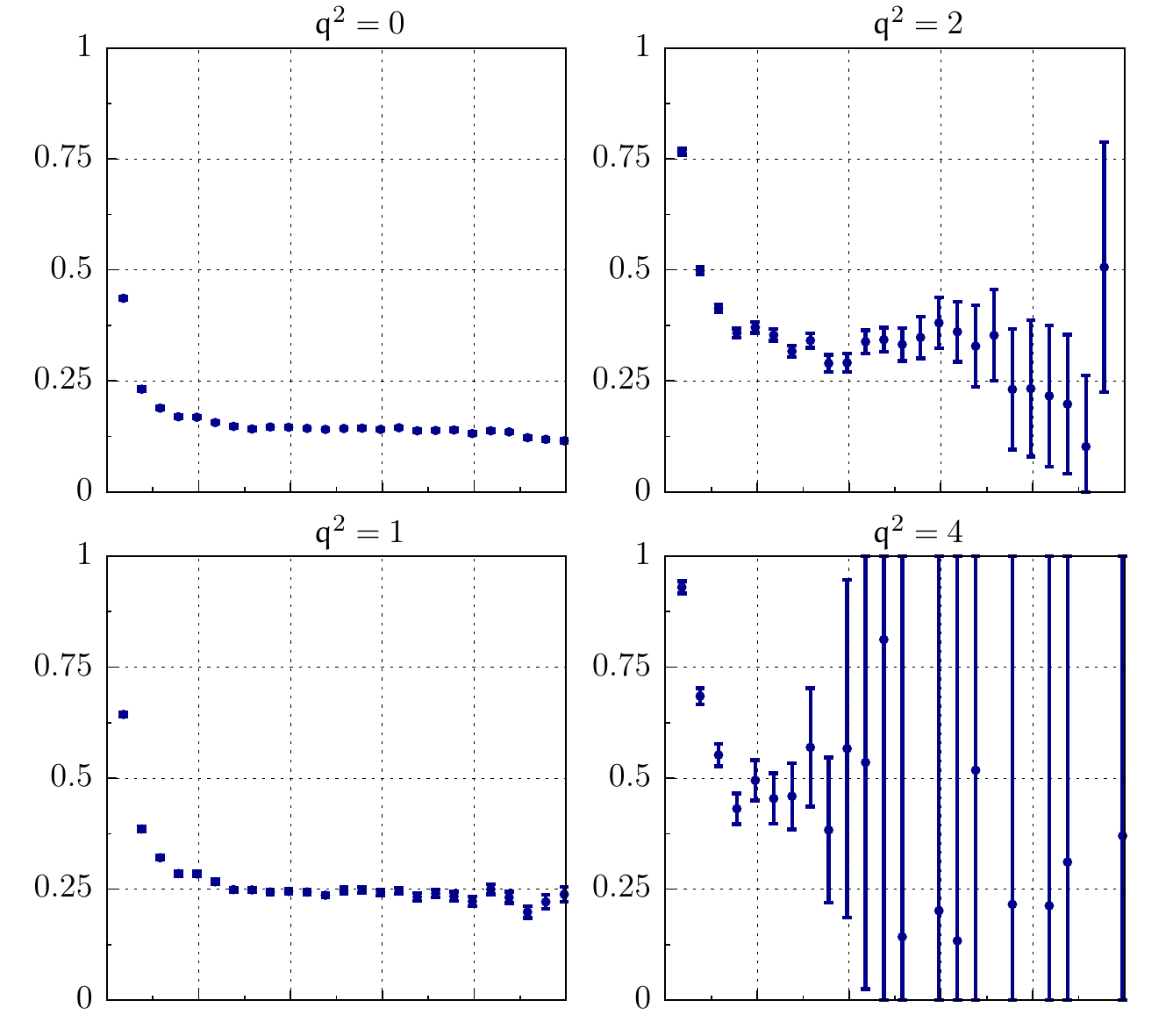}
\caption{\small Effective energies for pions at different values of ${\mathbf q}^2$}
\label{fig:RNS}
\end{figure}

In general, the problem becomes more severe as non-zero spatial momenta are considered, as for instance
in the computation of form-factors, and it is therefore worth
trying to optimise the smearing procedure depending on the kinematics. 
The most prominent example is given by the pion. At zero momentum the noise-to-signal ratio 
($R_{\rm NS}$) is expected
to be constant as a function of the time $x_0$, whilst for ${\mathbf{p}} \neq 0$, from a generalization of the arguments 
presented in~\cite{Parisi,Lepage} one expects
\begin{equation}
R_{\rm NS}(x_0) \propto e^{(\sqrt{m_\pi^2+{\mathbf p}^2}-m_\pi) x_0}\;, \qquad {\rm for} \; x_0\to \infty \;.
\label{eq:RNSasy}
\end{equation}
Throughout the paper we will characterise the momentum of the pion (nucleon) by a vector of integers, ${\mathbf q} = (q_1,q_2,q_3)$, where ${\mathbf p} = {\mathbf q} \,\frac{2\pi}{L}$.
In figure~\ref{fig:RNS}, we show the effective energies extracted from two-point correlation functions
of fields interpolating pions at non-zero spatial momentum ${\mathbf q}$ for different values of ${\mathbf q}^2$.
The plots refer to actual simulation results from measurements performed on an ensemble of $L/a=32$, 
$T/a=64$, $N_{\rm f}=2$ unquenched gauge configurations generated within the CLS initiative using 2 flavours of
non-perturbatively O($a$) improved Wilson fermions. Further numerical details can be found in~\cite{Fritzsch:2012wq,Capitani:2012gj}. 
In total, 168 configurations were used and for each of the configurations four evenly spaced source positions were utilised to 
improve the available statistics. We implemented Wuppertal smearing~\cite{Gusken:1989ad,Gusken:1989qx} at the source 
and sink with parameters $\kappa=2.9$ and $n=140$, whose meaning will be explained in the next section. These values were chosen to 
maximise the length of the effective mass plateau (i.e. ${\bf p}^2=0$) for pions and nucleons and produce a Gaussian wavefunction with a radius of 
approximately $0.5~$fm \cite{Capitani:2012gj}.

\begin{figure}[h!]
\begin{center}
\includegraphics[width=11.5cm]{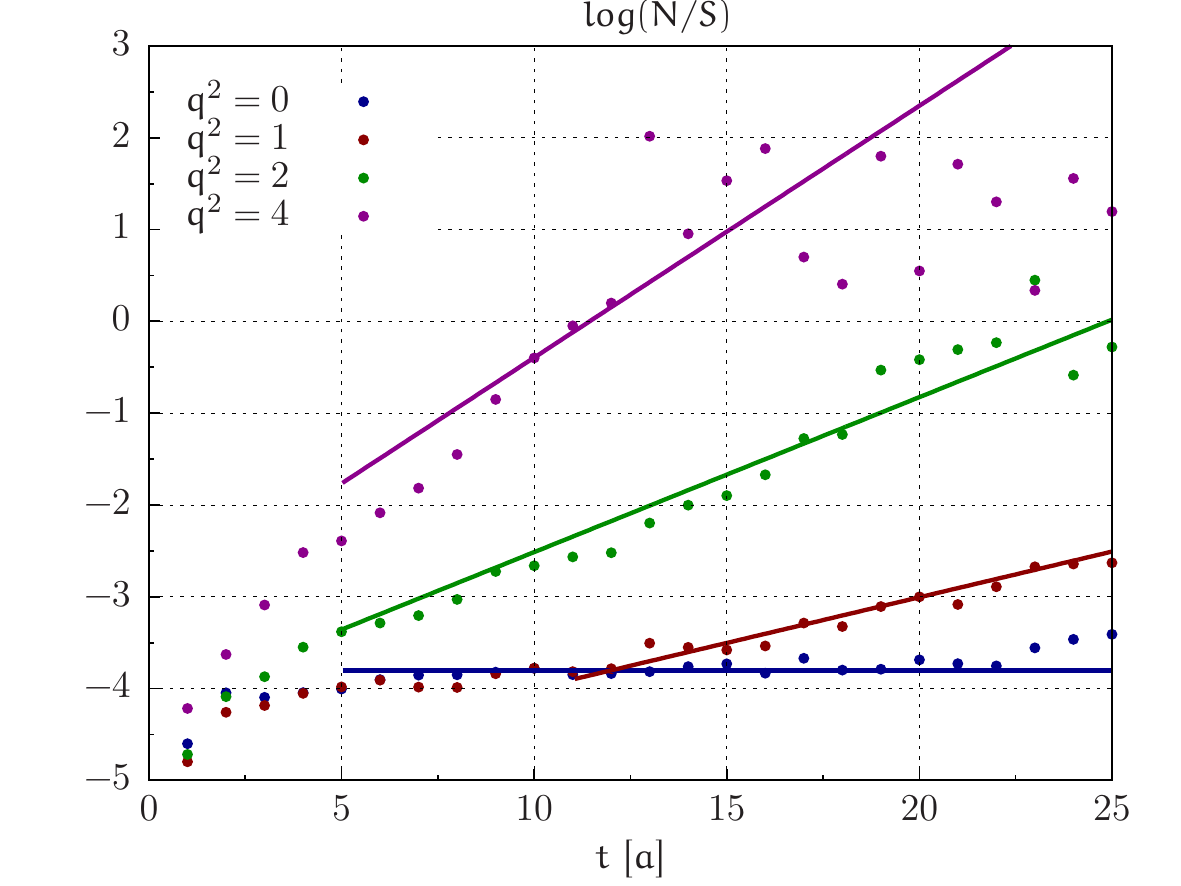}
\caption{\small Noise-to-signal ratio for different ${\mathbf q^2}$ plotted against the asymptotic behaviour of eq.~(\ref{eq:RNSasy})}
\label{fig:RNSasy}
\end{center}
\end{figure}

The dramatic deterioration of the signal as ${\mathbf q}^2$ increases is quite striking. 
In figure~\ref{fig:RNSasy}, the function $\log R_{\rm NS}(x_0)$ from the same dataset 
is plotted as a function of $x_0$ for different values of ${\mathbf{q}}^2$ and it is compared
to the expected asymptotic behaviour given in eq.~(\ref{eq:RNSasy}). The asymptotic 
trend sets in quite early in the example and it is consistent with
the noise being dominated by a zero-momentum two pion state.

In the following, we will show how a significant improvement can be obtained by adopting non-spherical quark-smearing
in order to interpolate boosted hadrons, instead of using the spherically symmetric function optimised for the
zero-momentum case, as it is usually done. We underline that whilst this method provides a reduction of the noise in the measured quantities (the effective energies in this study), it does not achieve an exponential improvement of the noise-to-signal ratio.

A preliminary account of our work has been given in~\cite{Poster}. Here we study both the pion and nucleon channels.
From the comparison it is clear that a much bigger gain in noise reduction for boosted pions can be achieved due to 
the fact that, for non-zero lattice momenta, the Lorentz factor $\gamma$ is much larger for pions than it is for 
nucleons.

\medskip
\section{Anisotropic smearing for mesonic and baryonic 2-point functions}

We start by recalling the construction of Gaussian/Wuppertal smearing~\cite{Gusken:1989ad,Gusken:1989qx}.
Given the hopping operator $H_k$ in direction $k$
\begin{equation}
H_k(x,x')= U_k(x) \delta_{x',x+\hat{k}}+U^\dagger_k(x-\hat{k}) \delta_{x',x-\hat{k}}\;,
\end{equation}
the smeared fermion field $\psi^{(\kappa)}$ is defined as
\begin{equation}
\psi^{(\kappa)}(x) = (1+\kappa H)^n\, \psi(x)\;,
\label{eq:Wupp1}
\end{equation}
with 
\begin{equation}
H=H_x + H_y + H_z \;.
\label{eq:genH}
\end{equation}
In our generalization, we simply promote $\kappa$  and $H$ to spatial-vectors $\boldsymbol{\kappa}$ and ${\mathbf H}$ with components 
$x$, $y$ and $z$ and define
\begin{equation}
\psi^{(\boldsymbol{\kappa})}(x) = (1+\boldsymbol{\kappa}\cdot {{\mathbf H}})^n\, \psi(x)\;.
\label{eq:Wupp2}
\end{equation}

In this study, we fix $n=140$ and smear the gauge links 
in the hopping  operator by applying one level
of hypercubic smearing~\cite{Hasenfratz:2001hp}.
For the particular choice of smearing parameters, we follow the definition of
``HYP2" links in~\cite{DellaMorte:2005yc}.
We will consider  two-point correlation functions of the type
\begin{equation}
C(x)= \left\langle O(\psi^{(\boldsymbol{\kappa})}(x),\overline{\psi}^{(\boldsymbol{\kappa})}(x)) \, 
O^\dagger(\psi^{(\boldsymbol{\kappa})}(0),\overline{\psi}^{(\boldsymbol{\kappa})}(0))\right\rangle \;,
\end{equation}
where the operator $O$ interpolates mesons or baryons and is therefore bi- or tri-linear
in the fields  $\psi^{(\boldsymbol{\kappa})}$ and $\overline{\psi}^{(\boldsymbol{\kappa})}$.
Our goal is to reduce the zero-momentum component, which dominates the noise, or 
equivalently to enlarge the high-momentum components in the four-point 
correlation function describing the variance. This is achieved by 
spatially squeezing the quark-wavefunction in the momentum direction. 
The choice is intuitively motivated by 
the physical picture that boosted hadrons are squeezed in the
momentum direction because of the relativistic space-contraction.
As long as the size of the resulting wavefunction
in the different directions is comparable to that
of a boosted hadron we do not expect to  spoil the good overlap 
with the ground state and therefore the fast approach to a plateau in the 
effective energy. 

\begin{figure}[h!]
\centering
\begin{minipage}[c]{0.5\linewidth}
    \includegraphics[width=1.0\linewidth]{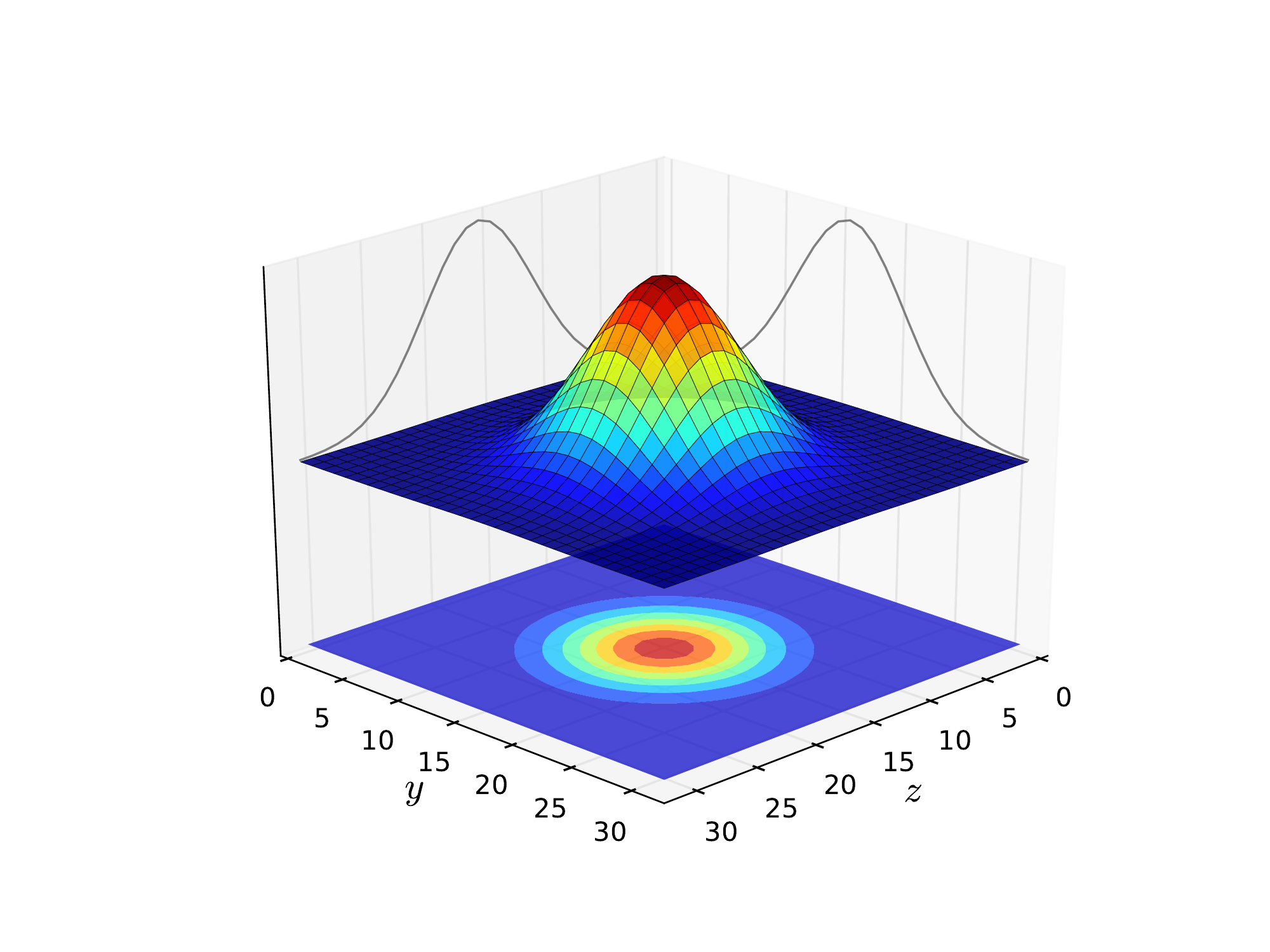}
  \end{minipage}
\hspace{-1cm}
  \begin{minipage}[c]{0.5\linewidth}
   \includegraphics[width=1.0\linewidth]{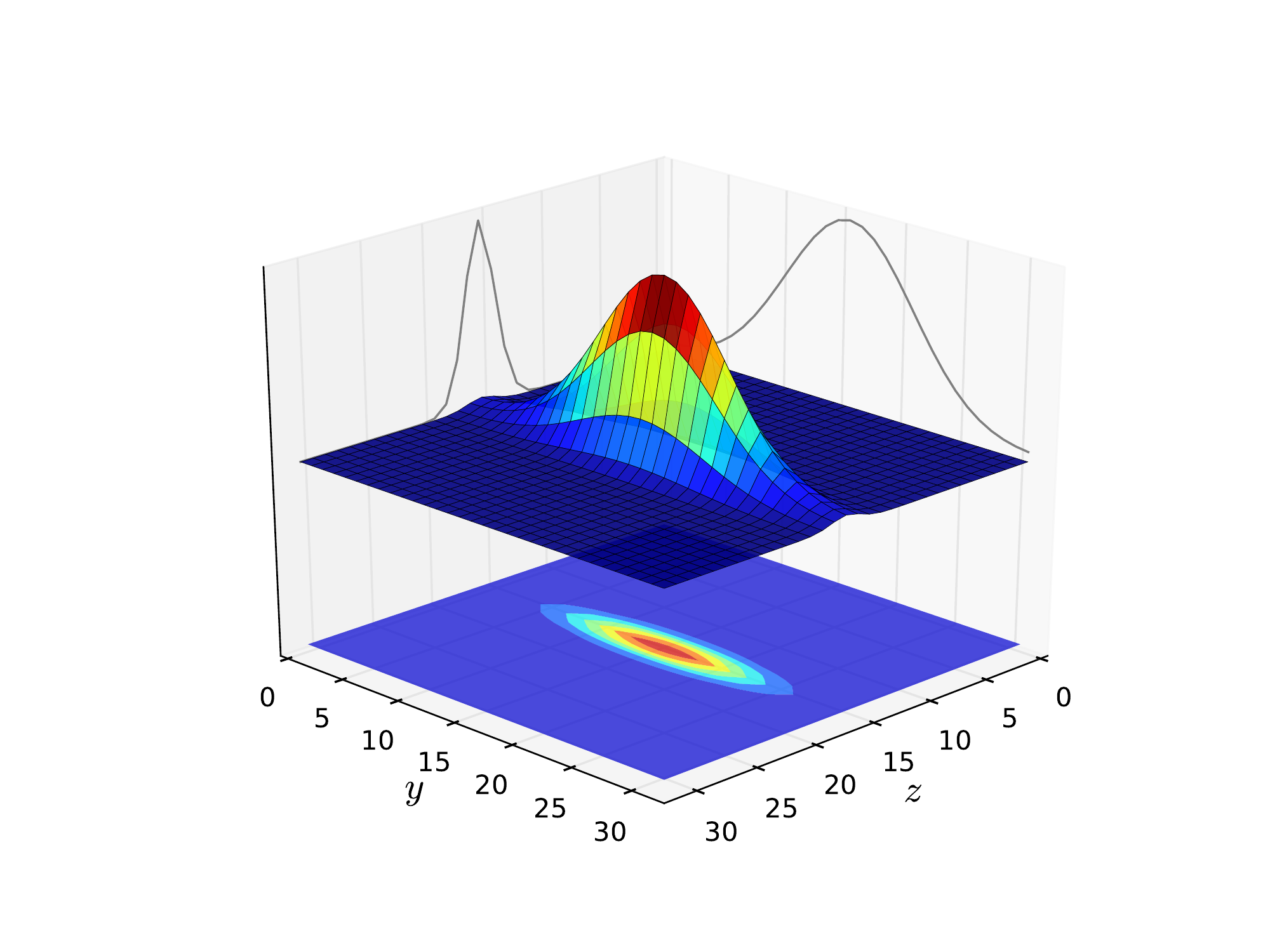}
  \end{minipage}
\caption{\small The $yz$-section of a spherical wavefunction $\kappa_x=\kappa_y=\kappa_z=2.9$ (left panel) and an anisotropic wavefunction $\kappa_x=\kappa_y=2.9,\kappa_z=2.9/16$ (right panel).}
\label{fig:wavefs}
\end{figure}

This generalization of Gaussian/Wuppertal smearing allows us to mix and
parametrically deform
point-like wavefunctions and broad Gaussian ones 
by tuning the components of $\boldsymbol{\kappa}$ in eq.~(\ref{eq:Wupp2}).
The first is known to give a better signal, especially at non-zero momentum,
whereas the latter results in a better plateau~\cite{Lin:2010fv}.
We show in figure~\ref{fig:wavefs}, the $yz$-section of a spherical wavefunction  
corresponding to $\kappa_x=\kappa_y=\kappa_z=2.9$ (left panel) as well as the $yz$-section of an anisotropic
one, obtained using $\kappa_x=\kappa_y=2.9,\kappa_z=2.9/16$ (right panel).
The lattice used is $32^3 \times 64$ points with a lattice spacing of $0.063(2)\,$fm and a pion mass
of roughly $450\,$MeV~\cite{Fritzsch:2012wq,Capitani:2011fg}.

\begin{figure}
\centering
\includegraphics[width=1.0\linewidth]{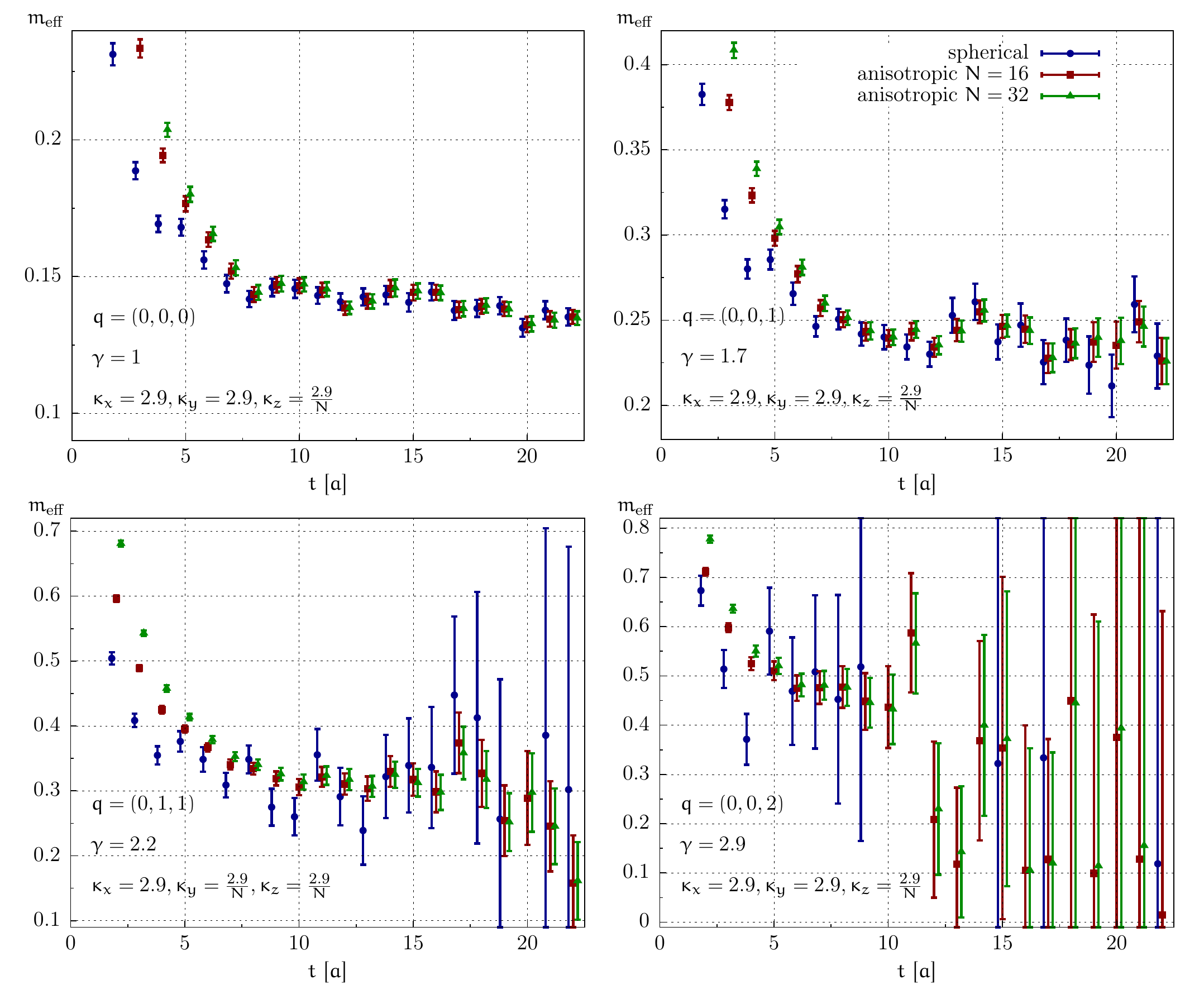}
\caption{\small Effective energies for the pion
 for different values of ${\bf q^2}$ using spherical (blue) and anisotropic 
(red and green) smearing.}
\label{fig:Meff_Pion}
\end{figure}

The anisotropic smearing is used so that the spatial squeezing is applied only to the directions 
with momenta. A numerical determination in one dimension demonstrated that for fixed $n$ in eq.~(\ref{eq:Wupp2}), 
the size of the wave function in direction $l$ depends on $\kappa_l$ as roughly $\kappa^{1/5}$.
Figure~\ref{fig:Meff_Pion} compares the effect of spherical and anisotropic smearing 
on the effective energies for the first four Fourier momenta of the pion. This demonstrates a clear 
gain in the statistical signal, whilst preserving a good overlap with the ground state. The gain is 
more pronounced for an increasingly boosted pion. For instance, for ${\bf q}=(0,1,1)$ and $(0,0,2)$ 
the identification of the plateau and the statistical error is significantly improved.
The plots display the direction of squeezing as well as the corresponding values of $\gamma$ to indicate 
the amount of boosting, where
\begin{equation}
\gamma = {{\sqrt{m_{\rm H}^2 +{\bf p}^2}}\over{m_{\rm H}}}
\label{eq:Relgam}
\end{equation}
for a hadron H.

The amount by which we spatially squeeze the directions with momenta should be dependent on the amount 
of boosting, $\gamma$. We checked this behaviour through implementing 
several choices for the wavefunction scaling $1/N$, with $N=2,~4,~8,~16$ and $32$. 
Figure~\ref{fig:Meff_Pion} demonstrates that the improvement in statistical noise has saturated by $N=16$ and that any further improvement for $N=32$ is accompanied by a slightly reduced suppression of excited states.
In order to further corroborate our intuitive argument we also checked that squeezing the $z$-direction for ${\bf q}=(1,0,0)$ or $(1,1,0)$, i.e.
squeezing in an orthogonal direction to the momentum, does not provide a gain.

\begin{figure}[h!]
\begin{center}
\includegraphics[width=11.5cm]{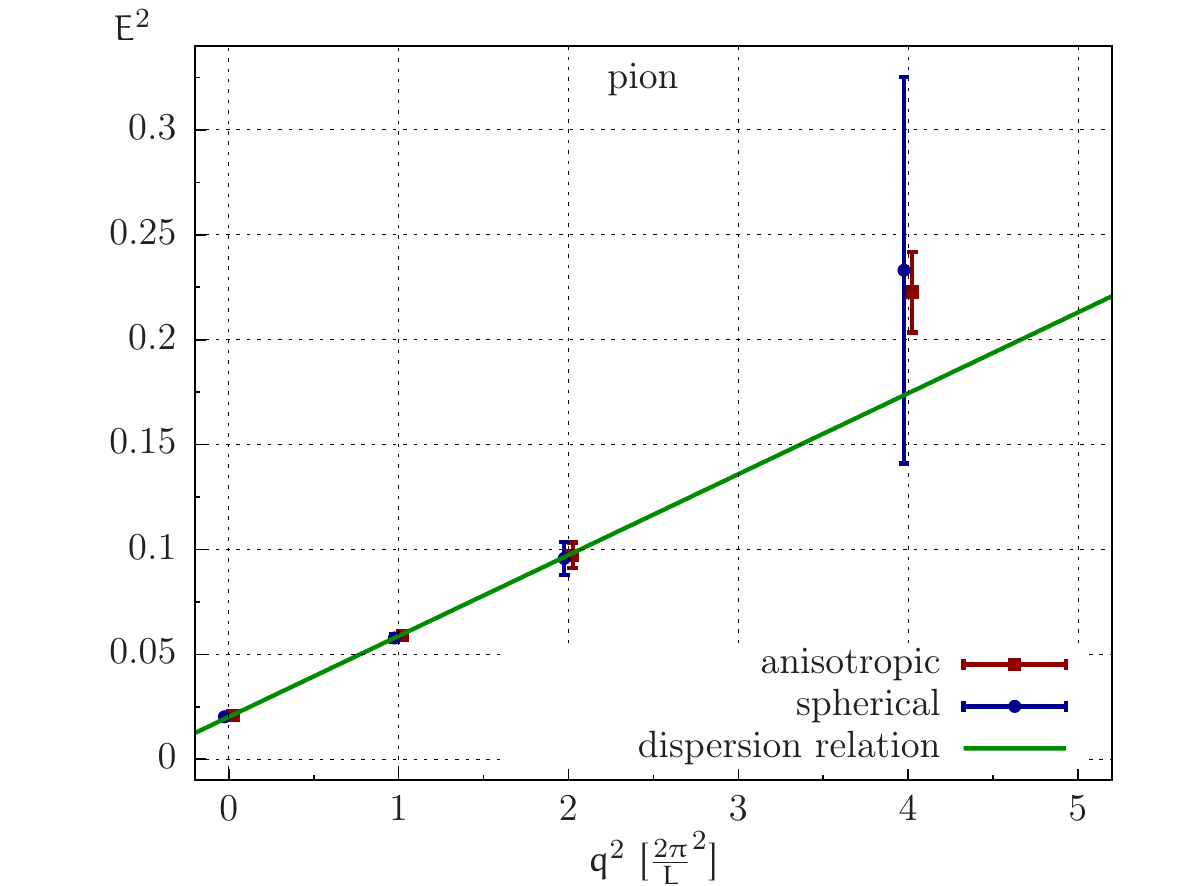}
\caption{\small The dispersion relation plotted against the extracted effective energies for spherical (blue) and anisotropic (red) 
smearing for the pion. The solid
    line denotes the continuum dispersion relation.}
\label{fig:Disp_Pion}
\end{center}
\end{figure}

The gain achieved by anisotropic smearing is further illustrated in figure~\ref{fig:Disp_Pion}, which shows the energies extracted from plateau fits
to figure~\ref{fig:Meff_Pion}. The fit-ranges are chosen to be the same for spherical and anisotropic smearing, and are determined through a $\chi^2/\textrm{dof}$ minimisation with a maximal plateau length.
The continuum dispersion relation, determined from the physical pion mass, is overlaid to indicate the expected behaviour. 

\medskip
%\section{Baryonic 2-point functions}

\begin{figure}[h!]
\centering
\includegraphics[width=1.0\linewidth]{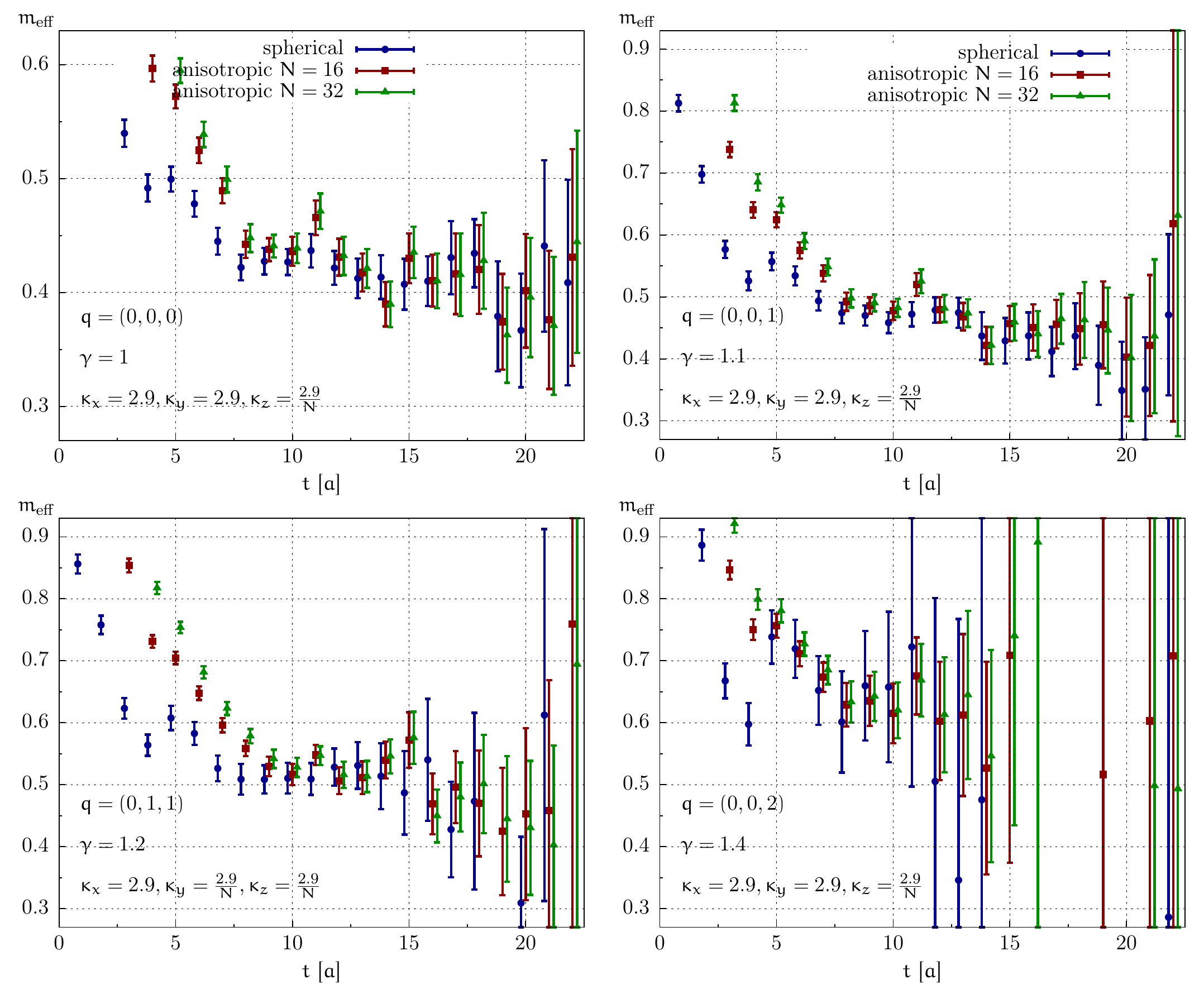}
\caption{\small Effective energies for the nucleon for different values of ${\bf q^2}$ using spherical (blue) and anisotropic (red and green) smearing.}
\label{fig:Meff_Nucl}
\end{figure}

\begin{figure}[h!]
\begin{center}
\includegraphics[width=11.5cm]{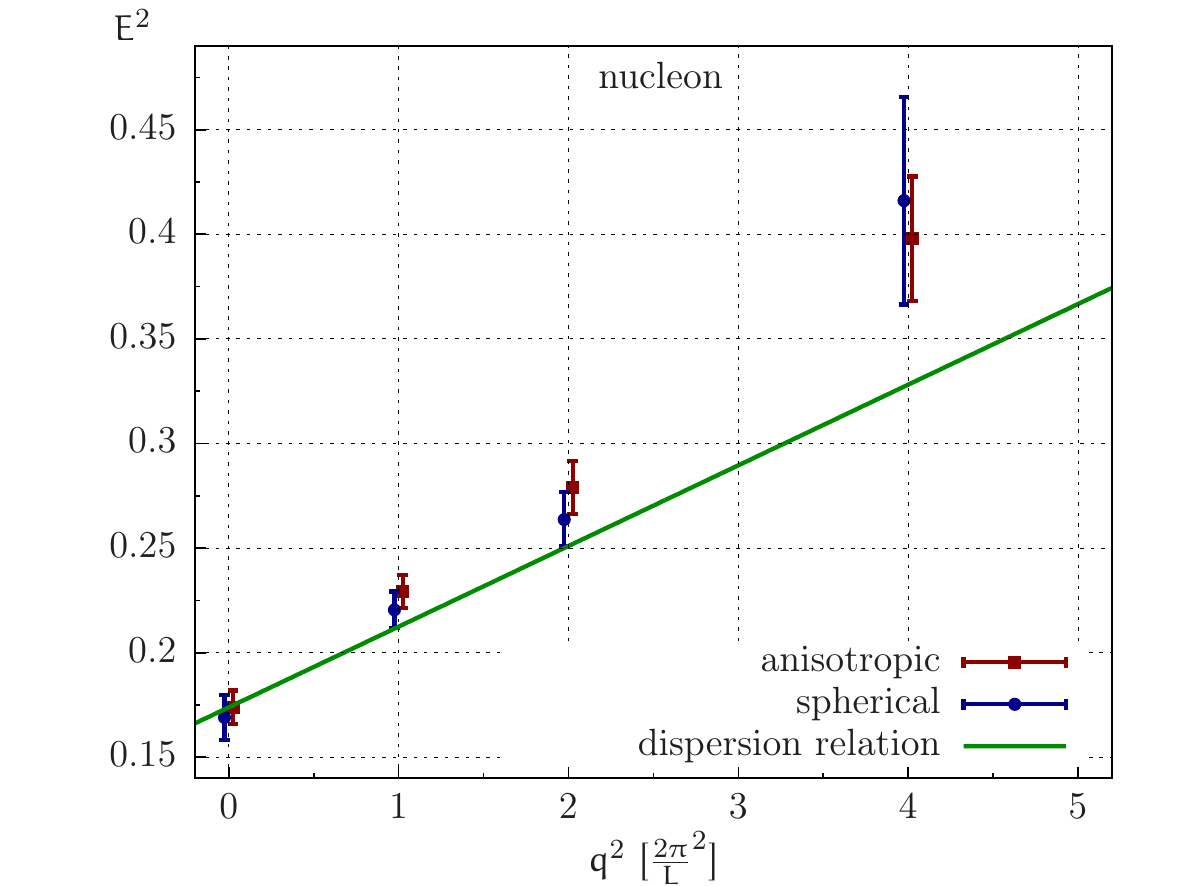}
\caption{\small The dispersion relation plotted against the extracted effective energies for spherical (blue) and anisotropic (red) smearing for the nucleon.}
\label{fig:Disp_Nucl}
\end{center}
\end{figure}

We apply the same treatment used for the pion to the nucleon, figures~\ref{fig:Meff_Nucl} \& \ref{fig:Disp_Nucl}. 
However, the signal-to-noise ratio is worse to begin with
and so we are restricted from looking at significantly boosted nucleons. This means $\gamma\sim 1$ and therefore the gain is not as striking as for the pion. The most likely explanation for the observed difference between the dispersion relation and the data points is the presence of cutoff effects. The latter can be estimated from a comparison between the continuum dispersion relation and a linear fit of the three data points with the smallest momenta for both the spherical and anisotropic smearing. The difference of 4\% amounts to less than two sigma and is therefore not statistically significant.

\medskip
\section{Conclusions and outlook}

We have provided numerical evidence that by a simple generalization of Gaussian/Wuppertal smearing, which allows us to produce non-spherical quark-wavefuctions,
the  noise-to-signal ratio in correlation functions describing the propagation
of hadrons at non-zero momentum can be significantly reduced.
The prescription yields a reduction of the statistical uncertainty without
affecting the onset of a plateau in the effective energy and consequently
the overlap of the wavefunction on the ground state.
The strongest evidence is for the case of pions, as those are already highly boosted
at small values of the typical lattice momenta. The comparison between pions and nucleons also indicates that the statistical gain mainly depends on the boosting factor, $\gamma$, rather than on the mass of the hadron. For this reason we believe this method to be even more effective as the chiral limit is approached. Of course, this expectation requires further studies, which we will consider in the future. 

Recently, a very similar 
approach has been proposed in~\cite{Roberts:2012tp} for 
nucleon interpolating fields. In that case, only a small improvement has been observed, which is consistent
with our findings.

The largest gain can probably be achieved by combining the approach with 
variational  techniques, as in the Generalized Eigenvalue Problem 
method~\cite{Luscher:1990ck,Blossier:2009kd}. As a different smearing
should in principle be adopted for each momentum, the use of 
all-to-all~\cite{Foley:2005ac,Endress:2011jc} stochastic propagators could be advantageous
in order to contain the numerical cost by reducing the number of inversions
of the Dirac operator to be performed.
We will explore these possibilities in future studies.

The main applications concern the determination of mesonic and baryonic form 
factors, e.g. for the $B \to \pi l \nu$ process or for the nucleon form 
factors $G_{\rm E}$ and $G_{\rm M}$, on which we have recently presented
preliminary results~\cite{talk}.

\medskip

\noindent{\bf{Acknowledgements:}} We are grateful for useful discussions with Bastian Brandt. Simulations were performed on the dedicated QCD platform ``Wilson" at the
Institute for Nuclear Physics, University of Mainz and on ``JUGENE" at Forschungszentrum J\"ulich. We are indebted to the
Institutes for their technical support and to the members of CLS for sharing
the configurations. This work was partially supported by  the Research Center EMG 
funded by {\it Forschungsinitiative Rheinland-Pfalz}. TR acknowledges support from DFG Grant $\textrm{HA4470/3-1}$

%\bibliography{biblist}           %or whatever

\begin{thebibliography}{100}
\bibliographystyle{JHEP}   %if you use h-elsevier.bst
%
\bibitem{Parisi}
G. Parisi, Phys. Rept. 103 (1984) 203.
%
\bibitem{Lepage}
G.P. Lepage, TASI 89 Summer School, Boulder, CO, Jun 4-30, 1989.
%
\bibitem{Fritzsch:2012wq}
  P.~Fritzsch, F.~Knechtli, B.~Leder, M.~Marinkovic, S.~Schaefer, R.~Sommer and F.~Virotta,
  %``The strange quark mass and Lambda parameter of two flavor QCD,''
    Nucl.\ Phys.\ B {\bf 865} (2012) 397 
    [arXiv:1205.5380 [hep-lat]].
%
\bibitem{Capitani:2012gj}
  S.~Capitani, M.~Della Morte, G.~von Hippel, B.~J\"ager, A.~J\"uttner, B.~Knippschild, H.~B.~Meyer and H.~Wittig,
  %``The nucleon axial charge from lattice QCD with controlled errors,''
  arXiv:1205.0180 [hep-lat].
%
\bibitem{Gusken:1989ad}
  S.~G\"usken, U.~Low, K.~H.~M\"utter, R.~Sommer, A.~Patel and K.~Schilling,
  %``Nonsinglet Axial Vector Couplings Of The Baryon Octet In Lattice Qcd,''
  Phys.\ Lett.\ B {\bf 227} (1989) 266.
%
\bibitem{Gusken:1989qx}
  S.~G\"usken,
  %``A Study of smearing techniques for hadron correlation functions,''
  Nucl.\ Phys.\ Proc.\ Suppl.\  {\bf 17} (1990) 361.
%
%\bibitem{Poster}
%  B.~J\"ager and T.~Rae, ``Improved interpolating fields for hadrons at non-zero momentum",
%poster presented at  "Lattice 2012", Cairns, Australia, June 26, 2012.
\bibitem{Poster}
  M.~Della Morte, B.~J\"ager, T.~Rae and H.~Wittig,
``Improved interpolating fields for hadrons at non-zero momentum",
 PoS LATTICE {\bf 2012} (2012) 260, 
 poster presented by BJ and TR at  "Lattice 2012", Cairns, Australia, June 26, 2012.
%
\bibitem{Hasenfratz:2001hp}
  A.~Hasenfratz and F.~Knechtli,
  %``Flavor symmetry and the static potential with hypercubic blocking,''
  Phys.\ Rev.\ D {\bf 64} (2001) 034504
  [hep-lat/0103029].
%
\bibitem{DellaMorte:2005yc}
  M.~Della Morte, A.~Shindler and R.~Sommer,
  %``On lattice actions for static quarks,''
  JHEP {\bf 0508} (2005) 051
  [hep-lat/0506008].
  %%CITATION = HEP-LAT/0506008;%%
%
\bibitem{Lin:2010fv}
  H.-W.~Lin, S.~D.~Cohen, R.~G.~Edwards, K.~Orginos and D.~G.~Richards,
  %``Lattice Calculations of Nucleon Electromagnetic Form Factors at Large Momentum Transfer,''
  arXiv:1005.0799 [hep-lat].
%
\bibitem{Capitani:2011fg}
  S.~Capitani, M.~Della Morte, G.~von Hippel, B.~Knippschild and H.~Wittig,
  %``Scale setting via the $\Omega\$ baryon mass,''
  PoS LATTICE {\bf 2011} (2011) 145
  [arXiv:1110.6365 [hep-lat]].
%
\bibitem{Roberts:2012tp}
  D.~S.~Roberts, W.~Kamleh, D.~B.~Leinweber, M.~S.~Mahbub and B.~J.~Menadue,
  %``Accessing High Momentum States In Lattice QCD,''
  arXiv:1206.5891 [hep-lat].
%
%\cite{Luscher:1990ck}
\bibitem{Luscher:1990ck}
  M.~L\"uscher and U.~Wolff,
  %``How To Calculate The Elastic Scattering Matrix In Two-dimensional Quantum Field Theories By Numerical Simulation,''
  Nucl.\ Phys.\ B {\bf 339} (1990) 222.
%
%\cite{Blossier:2009kd}
\bibitem{Blossier:2009kd}
  B.~Blossier, M.~Della Morte, G.~von Hippel, T.~Mendes and R.~Sommer,
  %``On the generalized eigenvalue method for energies and matrix elements in lattice field theory,''
  JHEP {\bf 0904} (2009) 094
  [arXiv:0902.1265 [hep-lat]].
%
%\cite{Foley:2005ac}
\bibitem{Foley:2005ac}
  J.~Foley, K.~Jimmy Juge, A.~O'Cais, M.~Peardon, S.~M.~Ryan and J.~-I.~Skullerud,
  %``Practical all-to-all propagators for lattice QCD,''
  Comput.\ Phys.\ Commun.\  {\bf 172} (2005) 145
  [hep-lat/0505023].
%
\bibitem{Endress:2011jc}
  E.~Endress, A.~J\"uttner and H.~Wittig,
  %``On the efficiency of stochastic volume sources for the determination of light meson masses,''
  arXiv:1111.5988 [hep-lat].
%
%\bibitem{talk}
%  T.~Rae, ``Excited state systematics in extracting nucleon
%electromagnetic form factors"
%  talk given at  "Lattice 2012", Cairns, Australia, June 28, 2012.
%
\bibitem{talk}
    S.~Capitani, M.~Della Morte, G.~von Hippel, B.~J\"ager, B.~Knippschild, H.~B.~Meyer and H.~Wittig,
 ``Excited state systematics in extracting nucleon electromagnetic form factors,''
 PoS LATTICE {\bf 2012} (2012) 177,
talk presented by TR at "Lattice 2012", Cairns, Australia, June 28, 2012.
%

\end{thebibliography}

\end{document}